\documentclass[twocolumn,showpacs,preprintnumbers,
amsmath,amssymb]{revtex4-1}
\usepackage{graphicx}

\def\beq{\begin{equation}}
\def\eeq{\end{equation}}
\def\bea{\begin{eqnarray}}
\def\eea{\end{eqnarray}}
\def\ar{\begin{array}}
\def\ear{\end{array}}

\def\nn{\nonumber}

\def\ga{\gamma}

\def\eps{\epsilon}

\def\rd{{\rm d}}

\def\rd{{\rm d}}

\makeatletter
\def\underbracket{%
  \@ifnextchar [ %
    {\@underbracket}%
    {\@underbracket [\@bracketheight]}}
\def\@underbracket[#1]{%
  \@ifnextchar [ %
    {\@under@bracket[#1]}%
    {\@under@bracket[#1][0.4em]}}
\def\@under@bracket[#1][#2]#3{
  \mathop {%
    \vtop {%
      \m@th \ialign {%
        ##\crcr $\hfil \displaystyle {#3}\hfil $%
e       \crcr \noalign %
       {\kern 3\p@ \nointerlineskip }%
        \upbracketfill {#1}{#2}
       \crcr \noalign %
       {\kern 3\p@ }%
     }%
   }%
  }%
  \limits%
}
\def\upbracketfill#1#2{%
  $\m@th \setbox \z@ \hbox {$\braceld$}
  \edef\@bracketheight{\the\ht\z@}\bracketend{#1}{#2}
  \leaders \vrule \@height #1 \@depth \z@ \hfill
  \leaders \vrule \@height #1 \@depth \z@ \hfill%
  \bracketend{#1}{#2}$%
}
\def\bracketend#1#2{\vrule height #2 width #1\relax}

\begin{document}
\title{Structures in the microwave background radiation}
\author{Krzysztof A. Meissner$^{1,2}$, Pawe\l~ Nurowski$^{1}$ and
  B\l a\.zej Ruszczycki$^{3}$}
\affiliation{$^1$ Faculty of Physics,
University of Warsaw,\\
Ho\.za 69, 00-681 Warsaw, Poland\\
$^2$ Max-Planck-Institut f\"ur Gravitationsphysik
(Albert-Einstein-Institut)\\
M\"uhlenberg 1, D-14476 Potsdam, Germany\\
$^3$ Nencki Institute, Polish Academy of Sciences, Warsaw, Poland\\
}
\begin{abstract} We compare the actual WMAP maps with artificial, purely
statistical maps of the same harmonic content to argue that there
are, with confidence level 99.7\%, ring-type structures in the observed cosmic microwave background.
\end{abstract}
\maketitle

\vspace{0.2cm}
\noindent{\bf 1. Introduction}

Since its discovery the cosmic  microwave background radiation (CMB)
has been the subject of very intensive research. The CMB anisotropy provides invaluable information  not only about the Universe in the epoch of last scattering but also about the earlier epochs. It follows from the lucky coincidence that before the last scattering the Universe was radiation dominated, and in the presence of radiation pressure it could not evolve and all the structures essentially remained frozen. Therefore although the temperature at the last scattering was rather low ($\sim$ 1 eV), searching for structures of different types in the CMB anisotropy can possibly reveal facts about physics at energies far above those presently accessible in accelerators ($\sim$ 10  TeV).

In this paper we compare the real CMB maps in 3
frequency subbands W1 -- W3 as measured by WMAP
\cite{WMAP} with artificially produced purely Gaussian maps of the
same harmonic spectrum, looking for possible ring-type
structures in the temperature distribution (see \cite{P} for a theoretical
motivation).
In this comparison we used the HEALPix code \cite{HE} to visualize and handle
the maps but we used our own code to produce artificial maps. We find 2 statistically significant ring-type
structures on the real maps (both happen to be on the southern
Galactic hemisphere) that have no analogs on the artificial maps. We
compare several characteristics of the real maps and the
artificial maps. For all of these characteristics the real maps are
very different from the artificially produced random maps. We applied a quantitative test to measure the difference and it allowed us to conclude that with confidence level exceeding 99.7\% the ring-type structures we find on the CMB real maps are not the result of a statistical fluctuation.

\vspace{0.1cm}
\noindent{\bf 2. Creation of artificial maps}

Our algorithm for obtaining the maps with artificial CMB temperature distribution was as follows:
\begin{enumerate}
\item
we choose a maximal multipole number $L$ (in the present paper we use $L=1000$)
\item
for each multipole number $l$, $0\leq l\leq L$, we choose a suitable number $N_l$, a positive integer, which is the number of spherical harmonics contributing to the temperature for this given $l$
\item
for all integers $k_l$, $1\leq k_l\leq N_l$, we assign a direction $(\theta_{k_l},\phi_{k_l})=(\arccos(\pi x_{k_l}),2\pi y_{k_l})$ where
$x_{k_l}$ and $y_{k_l}$ are random numbers from the uniform distribution in the interval $[0,1]$
\item
we take the coefficient $C_l$ from the known real WMAP spectrum, and define a function $T_l=T_l(\theta,\phi)$ on the sphere by:
\beq
T_l(\theta,\phi)=\sqrt{\tfrac{C_l}{N_l}}\sum_{k_l=1}^{N_l} P_l(\cos(\omega(\theta,\phi,\theta_{k_l},\phi_{k_l})))
\eeq
where
$\omega(\theta,\phi,\theta_{k_l},\phi_{k_l})$ is the spherical angle between a given point $(\theta,\phi)$ on the sphere and the direction $(\theta_{k_l},\phi_{k_l})$
\item
we repeat points 2)-4) for each $l$, $0\leq l\leq L$
\end{enumerate}
The final formula for the temperature on the artificial map is:
$$T(\theta,\phi)=\sum_{l=0}^L T_l(\theta,\phi).$$
It depends on our choice of the integers $L$ and the $N_l$'s. There is no preferred direction on the sky since for each $l$ we choose randomly $N_l$ directions $(\theta_{k_l},\phi_{k_l})$ (therefore this procedure has the advantage over averaging over $m$ in the standard approaches). If we now perform the usual harmonic analysis of this map we get a set of $C_l$'s that differ from the WMAP $C_l$'s used in the procedure but the difference can be made as small as we wish with the choice of sufficiently large $N_l$'s. The artificial maps produced in this way
are purely statistical, Gaussian and, depending on the chosen $N_l$, can reproduce any prescribed spectrum of $C_l$ with arbitrary accuracy.

Since the spectrum of $C_l$ obtained from the real maps is (especially for small $l$) not precisely known we have produced artificial maps along the average, the upper and the lower $C_l$ curve. For all of them we obtained essentially the same result.

\vspace{0.1cm}
\noindent{\bf 3. Looking for ring-type structures}

We have excluded from all the
maps the Milky Way belt ($\pm 0.4$ radians above and below the Galactic equator) to avoid introducing structures caused either by the Milky Way contribution or by the subtraction procedure usually applied to the CMB maps.

The procedure for looking for the ring-type structure for any given real or artificial map consisted in the following:
\begin{enumerate}
\item
a grid of points with HEALPix (k=6) parametrization spreading over the entire
sphere has been created
\item
a function $\Theta_{\eps,\ga_k}(\ga)$:
\begin{displaymath}
\ \ \Theta_{\eps,\ga_k}= \left\{ \begin{array}{ll}
-\eps(\cos(\ga_k-\eps)-\cos\ga_k)^{-1} & \textrm{\ $\ga_k-\eps<\ga<\ga_k$}\\
\eps(\cos(\ga_k)-\cos(\ga_k+\eps))^{-1} & \textrm{\ $\ga_k<\ga<\ga_k+\eps$}\\
0 & \textrm{\ otherwise}
\end{array} \right.
\end{displaymath}
was defined (this form is dictated by the area measure on the sphere and the condition that a constant temperature should give no contribution to the integral below)
\item
for $N$ ($N\approx 7000$) directions $(\theta^i,\phi^i)$ ($i=1,\ldots,N$) belonging to the grid, for all widths of rings $\eps$ ($\eps=0.02$, 0.04 or 0.08 radians) and for
all angular ``radii'' $\ga_k=0.07+k\cdot0.01$ ($k=1,\ldots,23$) an integral
\beq
I_{(\eps,\ga_k)}(\theta^i,\phi^i)=\int\rd\Omega'\, \Theta_{\eps,\ga_k}(\omega(\theta^i,\phi^i,\theta',\phi'))\,T(\theta',\phi')\nn
\eeq
was calculated ($\omega(\theta^i,\phi^i,\theta',\phi')$ is the spherical angle between $(\theta^i,\phi^i)$ and $(\theta',\phi')$). In this way for a given $\eps$ we get a set of $N\cdot 23\approx 160000$ values
$I_{(\eps,\ga_k)}(\theta^i,\phi^i)$.
\item
for a given $\eps$ we order these $N\cdot 23$ values of $I_{(\eps,\ga_k)}(\theta^i,\phi^i)$ (separately for positive and negative values), group them in a number $n$ ($n=1500$) of bins with centers $x_i$ and for all artificial maps we form the normalized cumulative distribution $F_m(x_i)$  ($m$ stands for the map number and $i$ stands for the bin number)
\item
at each $x_i$ we average over all $F_m(x_i)$ to arrive at the average distribution $F(x_i)$
\item
we use the procedure for comparison of cumulative distribution functions introduced in \cite{KM}: for all artificial and 3 real maps W1 -- W3 we calculate
\beq
A=-\frac{a}{N}\sum_{i=1}^{n}d_i\ln(1-F^a(x_i))
\eeq
where $a$ is a positive number (we used several numbers, all very large, much larger than $n$) and $d_i$ is the number of points in the $i$th bin. Then, separately for positive and negative distributions, we check the hypothesis that a given map belongs to the distribution $F(x)$ using the formula given in \cite{KM} -- the formula quoted there is suitable since it allows for quantitative comparison of distribution functions differing mostly at the tails (as is the case here).
\end{enumerate}

The shape of the function $\Theta_{\eps,\ga_k}(\theta^i,\phi^i)$ defines what we mean by the ring-type structure -- the maximal integral is when the temperature is of the same shape. Since we gather in the same set all values of $\ga_k$ the same direction can give several large values of the integral so these values can be  correlated. The division into smaller number of bins $n$ ($n\ll N\cdot 23$) is intended to partially compensate for these correlations.

\vspace{0.1cm}
\noindent{\bf 4. Results}

It is commonly taken for granted (with the notable exception of \cite{GP}) that the Cosmic Microwave Background is purely statistical being produced by the quantum fluctuations usually assumed to have taken place during inflation (as the solution in De Sitter space suggests). Therefore it was very unexpected for us to find significant differences (with confidence level 99.7\%) between the WMAP results and artificial maps (with the same statistical properties as WMAP) that we have created.  We believe that the procedure of creating artificial maps described above, being different from the one usually
used, is interesting on its own. It gives purely statistical gaussian
maps, and reproduces the prescribed harmonic content with the desired
accuracy.

We have produced many artificial maps (to generate an artificial map with our procedure is fast) but the procedure for looking for ring-type structure required the number of integrals for any given map to be very large ($N\cdot 23\approx 160000$) and the CPU time needed was the main reason to limit ourselves in the present paper to 100 artificial maps only.

The differences between real and artificial maps were both qualitative and quantitative. First of all, for $\eps=0.08$ on frequency
subband maps W1 -- W3 we find two directions around which there is a significant concentration of circular structures distinguished by the large values of $I_{(\eps,\ga_0)}(\theta,\phi)$. The Galactic coordinates of these two directions are approximately $(\tilde\theta_1,\tilde\phi_1)=(2.6,3.7)$ (which correlates with the so cold ``cold spot'' \cite{CP1,CP2,CP3})
and $(\tilde\theta_2,\tilde\phi_2)=(2.6,2.9)$ (this is new and has opposite values of the integral -- the highest values from this direction are shown on Fig. 2). We interpret these two directions as two centers of circular rings in the temperature distribution. The reason for
this interpretation is as follows: in a close vicinity (about $8\cdot 10^{-3}$ sr) of the first of these directions, there are about 80 centers of circular structures (and there are none outside). A similar situation occurs around the second
direction -- there we find about 40 centers in about the same solid
angle as in the first case. We also found many more potential centers of ring-type structures but since they were less significant statistically we do not include them in the analysis. It is perhaps worthwhile to note that the positions of both of the main two ring-type structures and the less significant ones correlate approximately with the positions of the structures found in \cite{GP}.

The quantitative comparison between the real W1 -- W3 maps with the artificial maps
regarding the existence of structures with $\eps=0.08$ yields the
following:

For the procedure described above we get for the real maps W1 -- W3 with $n=1500,\ a/n=2000$ the values $A=107.9,\ 133.5,\ 84.8$ respectively, while only 6 out of 100 artificial maps have $A > 1$ (the largest value among these 6 maps is $A\approx 52$, the other 5 have much smaller $A$). According to \cite{KM} the probability of $A$ bigger than $\sigma$ for large $a/n$ is approximately equal to ($1\ll \sigma\ll a/n$)
\beq
P(A>\sigma)=\frac{n}{a}\ln\frac{a}{n\sigma}
\eeq
so we get for the real map $W2$ the probability of a statistical fluctuation $\approx 0.3\%$, which means that with 99.7\% confidence level we can reject the hypothesis that the values of the integrals on the real maps are purely statistical and state that the ring type structures on the background radiation maps are not statistical fluctuations.
In favor of this statement we can also observe that for the value of $I_{(\eps,\ga_0)}(\theta,\phi)$ greater than 3 on the map of frequency band W2 only 81 points are in the northern hemisphere and 1430 in the southern one -- the difference is similar for two other spectral types but there are no such differences for the artificial maps.

It is interesting to note that for $\eps=0.02$ and $\eps=0.04$ the differences between the real and artificial maps are much smaller.

For comparison we have also imposed a
condition that around the same direction $(\theta,\phi)$ we have both
large positive and large negative integrals. It turned out that none of the real maps shows any statistically significant presence of such structures -- it points to the presence of one sharp circular edge in the temperature distribution and not two or more (which may be there but milder).

In figure 1 we have plotted the directions of the suspected ring-type structures
on the real WMAP maps in 3 frequency bands W1 -- W3.(with the galactic disk excluded, as described in the text).
In fig 2. we have plotted the upper part of the histogram of the cumulative distribution functions for all artificial maps and three real maps W1 -- W3 (for positive value of the integral).

\vspace{0.2cm}
\noindent{\bf {Acknowledgments:}} We gratefully acknowledge helpful discussions with Pawe{\l} Bielewicz, Marek
Demia\'nski, Krzysztof G\'orski and Roger Penrose.
Special thanks are due to  C. Denson Hill and E. Ted Newman for many
suggestions on preliminary versions of the article.

\onecolumngrid
\vspace{0.7cm}

\begin{center}
\includegraphics[width=12cm,viewport= 0 0 950 480,clip]{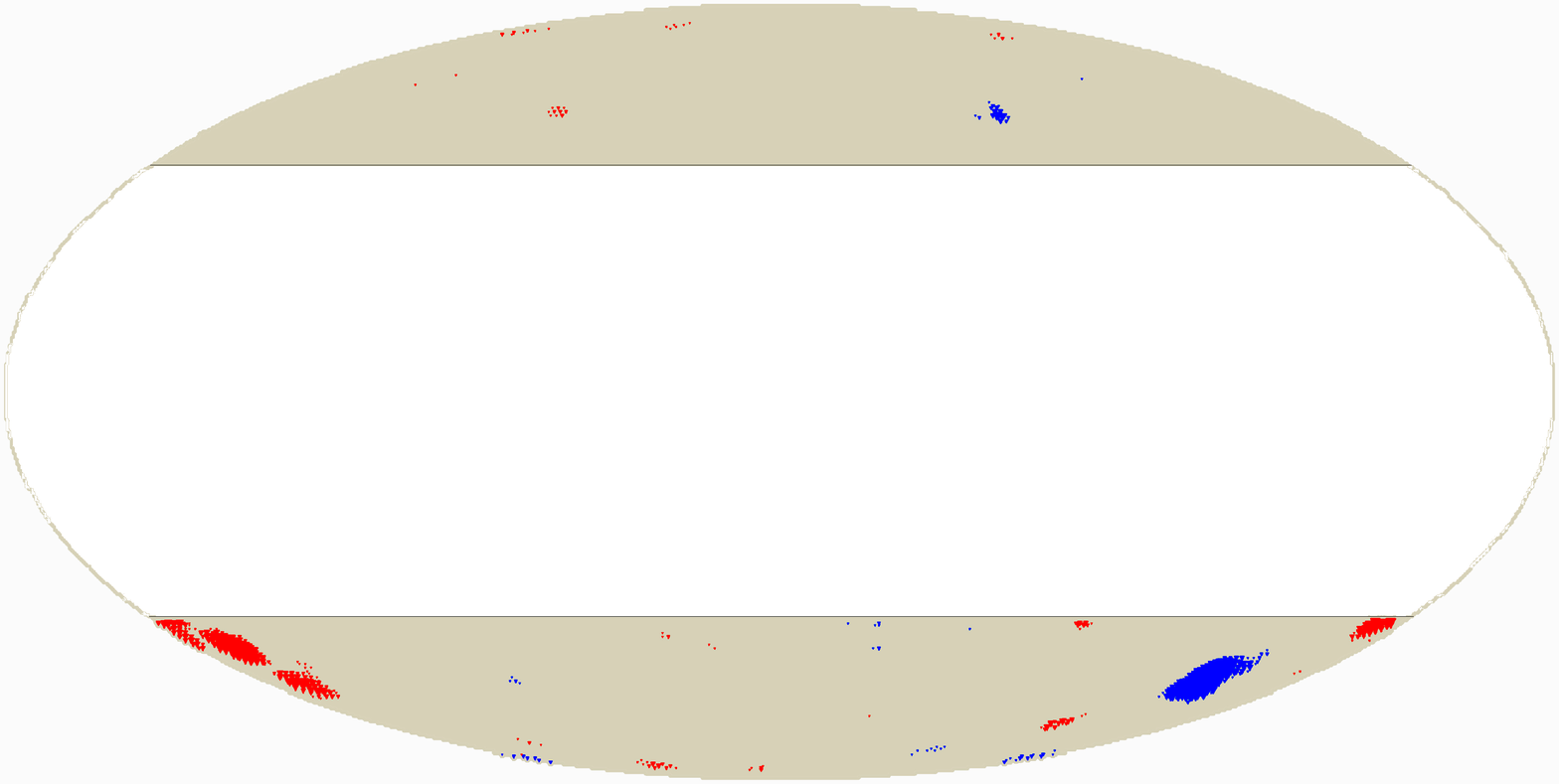}

\vspace{0.5cm}
Fig.1. Position of centers of circular structures on the real maps W1 -- W3
\end{center}

\vspace{0.1cm}

\onecolumngrid
\begin{center}
\includegraphics[width=10cm,viewport= 0 0 620 420,clip]{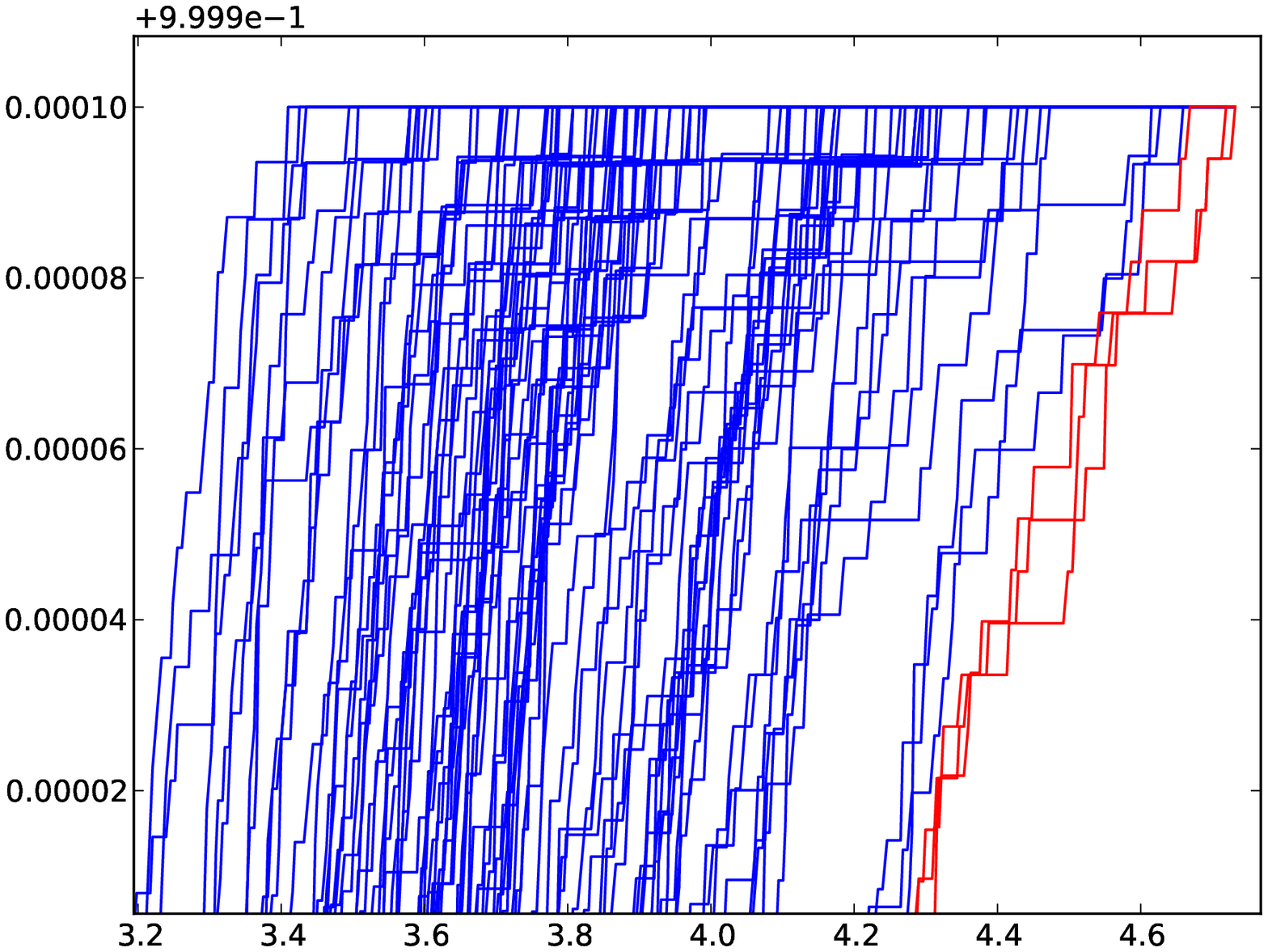}

\vspace{0.2cm}
Fig.2. The tail of the distributions of the integrals of the 100 artificial (blue) and 3 real (red) maps
\end{center}


\end{document}